\documentclass[pra,aps,twocolumn,floatfix,showpacs,tightenlines,amsmath,amssymb,superscriptaddress]{revtex4}

\usepackage{amsmath}
\usepackage{epsfig}

\usepackage{graphicx}
\usepackage{dcolumn}
\usepackage{bm}

\newcommand{\ket}[1]{|{#1}\rangle}

\begin{document}

\title{Quantum error correction via robust probe modes}

\author{Fumiko Yamaguchi}\email{yamaguchi@stanford.edu}
\affiliation{E. L. Ginzton Laboratory, Stanford University,
Stanford, CA 94305, USA}
\author{Kae Nemoto}\email{nemoto@nii.ac.jp}
\affiliation{ National Institute of Informatics, Hitotsubashi,
Chiyoda-ku, Tokyo 101-8430, Japan}
\author{William J. Munro}
\affiliation{Hewlett-Packard Laboratories, Filton Road, Stoke
Gifford, Bristol, BS34 8QZ, UK}

\date{\today}

\begin{abstract}
We propose a new scheme for quantum error correction using robust
continuous variable probe modes, rather than fragile ancilla
qubits, to detect errors without destroying data qubits. The use
of such probe modes reduces the required number of expensive
qubits in error correction and allows efficient encoding, error
detection and error correction. Moreover, the elimination of the
need for direct qubit interactions significantly simplifies the
construction of quantum circuits. We will illustrate how the
approach implements three existing quantum error correcting codes:
the 3-qubit bit-flip (phase-flip) code, the Shor code, and an
erasure code.
\end{abstract}


\maketitle

In recent years, we have seen the development and realization of
small scale quantum devices and circuits
\cite{Nielsen2000,lo98,gis02,spi05}. While these systems currently
use only a few qubits, they show great promise for being able to
use in large scale quantum computation. As the size of such
systems grows, the investigation of practical schemes for
fault-tolerant architectures has become an urgent task. The
standard method of error correction is to use several physical
qubits to encode quantum information in a redundant fashion and
then introduce ancilla qubits to perform error syndrome detection
and correction \cite{Shor1995,Steane1996,Calderbank1996}. The
concatenation of these error correction schemes has been shown to
guarantee the computational system to be fault tolerant for error
rates below a certain threshold value \cite{Shor1996}. Thus the
total system in principle satisfies efficiency and scalability.
However, in practice, implementation of a system for
fault-tolerant quantum computation is a daunting task due to the
huge overhead in number of qubits and fundamental quantum logic
gates.

In the standard method of error correction, ancilla qubits are
used to project the state of the logical qubit onto one of several
subspaces, and thus the task of ancilla qubits is rather simple.
Hence it might be possible to replace the costly ancilla qubits
with something more practically viable such as continuous-variable
probe modes. In this paper, we present a new scheme to implement
existing error correcting codes, replacing the ancilla qubits with
robust probe modes. The introduction of probe modes interacting
nonlinearly with qubits facilitates the encoding, error detection
and correction procedures. The interaction between qubits and a
probe mode enables detection of errors in the qubits by
measurement of the probe mode. Such operations can be achieved in
variety of physical systems \cite{Spiller2005}. For example, for
electron spin qubits, the interaction can be achieved with a
coherent light interacting with the electrons contained in optical
microcavities
\cite{Santori2002,Reithmaier2004,Yoshie2004,Akahane2003}. In the
all-optical implementation, such operations can be achieved with
an intense coherent light interacting with photonic qubits via a
cross-Kerr nonlinearity \cite{Nemoto2004,Munro2005A,Munro2005B}.
In these physical systems, the probe mode is particularly stable
in its preparation \footnote{We refer a reader to a recent review
by S. L. Braunstein and P. van Loock, Rev. Mod. Phys. {\bf 77},
513 (2005), for a concise introduction to quantum information with
continuous variables.}, and this provides flexibility in arranging
qubits in space by connecting them remotely. To illustrate our
approach, we demonstrate implementations of the following codes:
(1) the 3-qubit bit (or phase) flip code \cite{Nielsen2000}, (2)
the 9-qubit Shor code \cite{Shor1995} and lastly (3) an erasure
code \cite{Ralph2005,Gilchrist2005}.

The essential component used in our approach is a weak nonlinear
interaction between a qubit and a probe mode (indicated by $p$).
This interaction is described by the Hamiltonian
\begin{equation}\label{phase shift}
    H =\hbar\chi |1\rangle\langle 1| \otimes {\hat n}_p,
\end{equation}
where $\chi$ is the interaction strength, ${\hat n}_p$ is the
number operator of the probe mode, and the basis states of the
qubit are given by $|0\rangle$ and $|1\rangle$. These basis states
could be, for example, the spin states of an electron, or the
polarization states of a photon. The process causes a conditional
phase shift on the probe mode dependent on the state of the qubit,
namely,
\begin{eqnarray} \label{nonlinear gate} \ket{0}\ket{\alpha}
\to \ket{0}\ket{\alpha}, \; \ket{1}\ket{\alpha} \to
\ket{1}\ket{\alpha e^{i\theta}}.
\end{eqnarray}
The size of the phase shift is given by $\theta = \chi t$ with $t$
being the interaction time. Furthermore, a significant advantage of
this scheme is that a probe mode can interact with multiple
qubits subsequently, each causing a phase shift in
the probe mode. Measurement of the accumulated phase shift of the
probe mode after the successive interactions allows the direct
measurement of collective properties of a multi-qubit state (more than
two) without destroying it. We will describe below two fundamental two-qubit gates based
on this elementary operation.

\begin{figure}[!htb]
\includegraphics[scale=0.4]{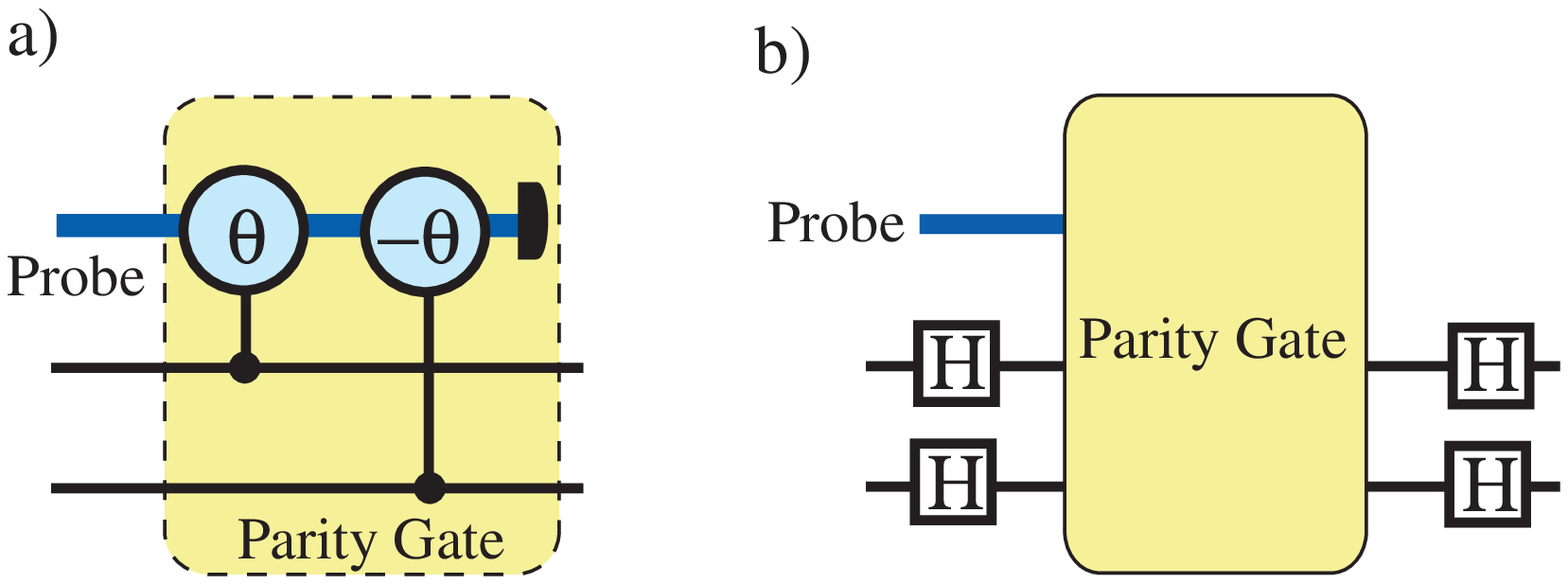}
\caption{(color online) a) A two-qubit parity gate. The
conditional phase shift given by the Hamiltonian Eq.~(\ref{phase
shift}) is denoted by $\theta$. After the interactions, the probe
mode is measured to project the qubit state onto the even or odd
parity subspace. b) A symmetrizer gate. The two-qubit parity gate
is applied in $\ket{+}$/$\ket{-}$ basis instead of
$\ket{0}$/$\ket{1}$ basis, and the second qubit is flipped when
the measured parity is odd.} \label{two-qubit parity gate}
\end{figure}

Our first fundamental gate, the two-qubit parity gate, is designed
to measure the product of Pauli operators $Z_1Z_2$ on qubits 1 and
2, and has been shown to be as useful as the controlled-NOT gate
for universal quantum computation \cite{Nemoto2005}. As shown in
Fig.~\ref{two-qubit parity gate}a), two qubits subsequently
interact with the probe mode, and the probe mode gains a phase
shift $+\theta$ when qubit 1 is in $\ket{1}$ and $-\theta$ when
qubit 2 is in $\ket{1}$. A two-qubit state with even parity
($\ket{00},\ket{11}$) causes no overall phase shift in the probe
mode, whereas one with odd parity ($\ket{01},\ket{10}$) causes a
phase shift $\pm \theta$ as in
\begin{equation}
\left\{\begin{array}{l}
\ket{00}\ket{\alpha} \to \ket{00}\ket{\alpha} \\
\ket{11}\ket{\alpha} \to \ket{11}\ket{\alpha}
 \end{array}\right., \;
\left\{\begin{array}{l}
 \ket{01}\ket{\alpha} \to \ket{01}\ket{\alpha e^{-i \theta}}\\
\ket{10}\ket{\alpha} \to \ket{10}\ket{\alpha e^{i \theta}}
 \end{array}\right..
 \end{equation}
A projective measurement on the probe mode distinguishes
$\ket{\alpha}$ and $\ket{\alpha e^{\pm i\theta}}$ (without
distinguishing $\ket{\alpha e^{i\theta}}$ from $\ket{\alpha
e^{-i\theta}}$). This projects the two-qubit state onto either the
even parity subspace (\{$\ket{00}$,$\ket{11}$\}) or the odd parity
subspace (\{$\ket{01}$,$\ket{10}$\})\cite{Munro2005A}. The
projection measurement can be achieved in a number of ways,
ranging from a straightforward $X$ homodyne measurement
\cite{Nemoto2004} to a photon number measurement. In this case,
the probe beam is displaced by an amount $-\alpha$ and then
measured with a QND photon number measurement \cite{Munro2005A}.
The even parity components result from a zero photon number
measurement, while the odd parity components arise from a non-zero
result. The odd-parity state can be converted to an even-parity
state by a classical feed-forward after the parity measurement. In
the construction of the parity gate, the direct interaction
between the qubits is not needed. Therefore, the gate can be
performed among distributed qubits.

The second fundamental gate is to create a two-qubit state with
$X_1X_2=+1$, starting with an arbitrary two-qubit state
$\ket{xy}$, where $x,y =0,1$. The gate generates the symmetrized
state, $(\ket{xy}+\ket{\bar{x}\bar{y}})/\sqrt{2}$. Hence we call
this gate the ``symmetrizer gate." (This gate was originally
introduced as the ``entangler gate" in \cite{Nemoto2004}.)
The symmetrizer gate corresponds to the two-qubit parity gate in
the $\ket{+}/\ket{-}$ basis and classical forward to flip the
second qubit if the measured parity is odd. The gate operation is
described by the basis change by the Hadmard transform $H$, the
two-qubit parity gate in the $\ket{0}/\ket{1}$ basis and another
$H$, as shown in Fig.~\ref{two-qubit parity gate}b).
The two-qubit state $\ket{xy}$ is written as
\begin{eqnarray}
    \ket{xy}&=&\frac{1}{2}\left[\ket{++}+(-1)^{x+y}\ket{--}\right]\nonumber\\
    &+&\frac{1}{2}(-1)^{x}
    \left[\ket{+-}+(-1)^{x+y}\ket{-+}\right],
    \label{xy_in_+-basis}
\end{eqnarray}
in the $\ket{+}$/$\ket{-}$ basis, where we have used $\ket{x} =
(\ket{+} + (-1)^{x}\ket{-})/\sqrt{2}$ and $\ket{y} = (\ket{+}
+(-1)^{y}\ket{-})/\sqrt{2}$. Then the two-qubit parity gate in the
$\ket{+}$/$\ket{-}$ basis and classical feed-forward to flip qubit
2, if the measured parity is odd, bring the two-qubit state to the
even state $(\ket{++}+(-1)^{x+y}\ket{--})/\sqrt{2}$, which is
 $(\ket{xy}+\ket{\bar{x}\bar{y}})/\sqrt{2}$ in the $\ket{0}$/$\ket{1}$ basis. This shows
the operation of the symmetrizer gate and indicates its potential
in the encoding procedures.

We next consider our error correcting scheme. An arbitrary unitary
operation on a qubit can be described as a linear superposition of
a bit-flip error ($X$), a phase-flip error ($Z$) and both. Thus,
unless a qubit is lost, an arbitrary quantum error is either a
bit-flip, a phase-flip or both \cite{Nielsen2000}. We first
describe the 3-qubit bit-flip code, which can detect and correct
at most one bit-flip error by utilizing redundant encoding. The
computational basis states for the logical qubit, $\ket{0}_L$ and
$\ket{1}_L$, are
\begin{equation}
\ket{0}_L=\ket{0}_1\ket{0}_2\ket{0}_3 \equiv \ket{000},
\ket{1}_L=\ket{1}_1\ket{1}_2\ket{1}_3 \equiv \ket{111}.
\end{equation}
The standard method of detecting a bit-flip error is to measure
the products of Pauli operators, $Z_1Z_2$ and $Z_2Z_3$, with the
help of two ancilla qubits and four controlled-NOT operations to
the ancilla qubits. The operators $Z_1Z_2$ and $Z_2Z_3$ constitute
the stabilizer of the code. The code subspace consists of their
simultaneous eigenstates with eigenvalue +1 ($\ket{000}$,
$\ket{111}$). If a bit-flip error occurs, one or both of the
stabilizer generators $Z_1Z_2$ and $Z_2Z_3$ becomes $-1$, as
summarized in Table~\ref{syndrome}. Measurement of each stabilizer
generator by this error correcting code is no more than a
two-qubit parity measurement. Therefore, the simple two-qubit
parity gate alone allows us error syndrome measurements, without
the help of ancilla qubits.
\begin{table}[htbp]
    \begin{tabular}{lccccc}
        Bit-flip error & \vline& $Z_1Z_2$ & $Z_2Z_3$ & \vline & modulo(4)\\
\hline
        None & \vline & $+1$ &  $+1$ & \vline & 0\\
        on qubit 1 & \vline& $-1$ & $+1$ & \vline & 2\\
        on qubit 2 & \vline& $-1$ & $-1$ & \vline & 3\\
        on qubit 3 & \vline& $+1$ & $-1$ & \vline & 1
    \end{tabular}
\caption{Error syndromes both in binary readout and in modulo(4)
readout.  The binary readout is given by the measurements of
$Z_1Z_2$ and $Z_2Z_3$ using two probe modes as shown in
Fig.~\ref{bit-flip-ghz}a), while the module(4) readout is given
using one probe mode by the circuit shown in
Fig.~\ref{bit-flip-ghz}b).} \label{syndrome}
\end{table}

\begin{figure}[!htb]
\includegraphics[scale=0.4]{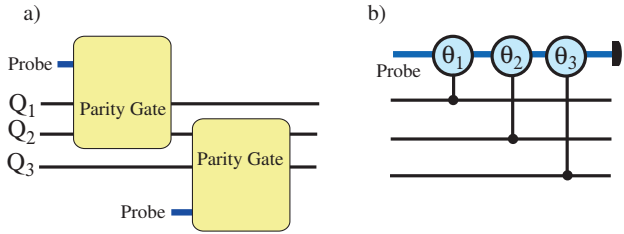}
\caption{(color online) Circuit for error detection by the 3-qubit
bit-flip code. In a) the first parity gate measures $Z_1Z_2$ and
the second parity gate measures $Z_2Z_3$. In b) we use one probe
beam but with the phase shifts ($\theta_1$, $\theta_2$ and
$\theta_3$) by qubits 1, 2, and 3. These are all distinct and
satisfy $\theta_1+\theta_2+\theta_3=0$. This gate works as a
modulo(4) error detection so as to give the readouts described in
Table~\ref{syndrome}.} \label{bit-flip-ghz}
\end{figure}

A circuit for the syndrome measurement gate (to measure $Z_1Z_2$
and $Z_2Z_3$) can be constructed as shown in
Fig.~\ref{bit-flip-ghz}a). If $Z_1Z_2$ is $+1$ ($-1$), the first
parity gate, involving qubits 1 and 2, measures even (odd) parity.
The operator $Z_2Z_3$ can be measured by the second parity gate
involving qubits 2 and 3. Therefore those two parity gates
complete error detection. Then the error can be corrected after
the detection by classical feed-forward control, or can be
recorded for later error correction.

Alternatively, by exploiting the novel benefits of the
probe-based scheme the measurements of $Z_1Z_2$ and $Z_2Z_3$ can be
simplified and performed by using only one probe mode instead of
two, as shown in Fig.~\ref{bit-flip-ghz}b). The interactions of
the probe mode $\ket{\alpha}$ with the three qubits are set such
that the phase shifts caused by the qubits 1, 2, and 3 are
$\theta_1$, $\theta_2$ and $\theta_3$, where $\theta_1$,
$\theta_2$ and $\theta_3$ are all distinct and satisfy
$\theta_1+\theta_2+\theta_3=0$ (for example, $\theta_1=\theta$,
$\theta_2=2\theta$, $\theta_3=-3\theta$). The encoded states
($\ket{000}$ or $\ket{111}$) cause no phase shift in the probe
mode. If a bit-flip error occurs on qubit 1 ($|100\rangle$ or
$|011\rangle$), the probe mode $|\alpha\rangle$ evolves into
$|\alpha e^{\pm i\theta_1}\rangle$. The probe mode gains a phase
shift $\pm\theta_2$ or $\pm\theta_3$ in the case of a bit-flip
error on qubit 2 or 3. By measuring the phase of the probe mode,
distinguishing $\ket{\alpha}$, $\ket{\alpha e^{\pm i\theta_1}}$,
$\ket{\alpha e^{\pm i\theta_2}}$ and $\ket{\alpha e^{\pm
i\theta_3}}$, we can identify an error. Those four error syndromes
correspond to different values in modulo(4) in
Table~\ref{syndrome}. This modification is equivalent to
replacement of two binary readouts by one modulo(4) readout, and
allows the use of the larger Hilbert space of the probe mode. As a
result, the error correcting procedure is significantly
simplified.

We now consider the encoding procedure.  The conventional circuit
to encode a unknown qubit state $c_0\ket{0} + c_1\ket{1}$,
initially stored in qubit 1, by the 3-qubit bit-flip code consists
of two controlled-NOT gates. Instead, we simply use the error
correcting procedure. To do so, we first initialize qubits 2 and 3
in $\ket{+}$. The three-qubit state is a sum of the four states
corresponding to the four error syndromes shown in
Table~\ref{syndrome}, and hence by running the error correcting
procedure described above we can reduce the state to
$c_0\ket{000}+c_1\ket{111}$. The one slight difference from the
error correcting procedure is that there should be no error on
qubit 1 in the encoding procedure. When the syndrome measurement
indicates an error on qubit 1, the result needs to be interpreted
as errors on qubits 2 and 3 and the classical feed forward has to
correct those two errors. The 3-qubit bit-flip code can be
converted to a code to correct at most one phase-flip error by
changing its basis from $\ket{0}$/$\ket{1}$ to
$\ket{+}$/$\ket{-}$.

The scheme for the 3-qubit error correcting code above can be
extended for the 9-qubit Shor code \cite{Shor1995}. This code is
constructed by concatenating the 3-qubit bit-flip code into the
3-qubit phase-flip code, and can correct an arbitrary error:
bit-flip error, phase-flip error or both. With the logical basis
states defined by $\ket{0}_L =(\ket{000}+\ket{111})^{\otimes
3}/2\sqrt{2}$ and $\ket{1}_L=(\ket{000}-\ket{111})^{\otimes
3}/2\sqrt{2}$ we can create the encoded state
$c_0\ket{0}_L+c_1\ket{1}_L$ from qubit 1, $c_0\ket{0}+c_1\ket{1}$,
as follows. We first initialize qubits 2 to 9 in $\ket{+}$, and
apply the 3-qubit bit-flip code on qubits 1/4/7, followed by the
Hadamard transformation $H$ on each of the qubits. This prepares
the state $c_0\ket{+}_1\ket{+}_4\ket{+}_7+
c_1\ket{-}_1\ket{-}_4\ket{-}_7$. Then we apply the 3-qubit
bit-flip code on each of the three sets of qubits 1/2/3, 4/5/6,
and 7/8/9. Such an operation leads to the encoded state
$c_0\ket{0}_L+ c_1\ket{1}_L$.

Next, in order to detect and correct errors we need to perform the
above procedure in roughly the reverse order. First we detect and
correct a bit-flip error if present. The 3-qubit bit-flip code is
applied to each of the three sets of qubits, 1/2/3, 4/5/6, and
7/8/9, and any errors are then corrected. Following this, we deal
with phase-flip errors. The phase-flip error correction requires
an additional step compared with the bit-flip error correction. We
first apply the symmetrizer gate on each of the pairs of qubits,
2/3, 5/6, and 8/9. The symmetrizer gate performs the following
action,
\begin{equation}
\ket{000}\pm\ket{111} \to \ket{\pm}(\ket{00}+\ket{11}),
\end{equation}
and thus disentangles qubits 2/3, 5/6, and 8/9 from qubits 1/4/7.
We now apply the 3-qubit phase-flip code on the remaining qubits
1/4/7 to correct phase-flip errors that may have occurred. After
the error detection and correction, we then need to re-encode our
quantum state. This can be done as described in the encoding
procedure.

The last error correcting code to be considered here will recover
an error caused by the complete loss of a qubit at a known
location, also known as qubit leakage.  This is the
dominant error in optical implementations, but can occur in most
physical implementations.  In some instances, this just
corresponds to the physical system leaving the qubit space.
Mathematically qubit loss corresponds to tracing out the qubit so
the information the qubit carried is completely lost, and thus the
density matrix of the remaining qubits becomes mixed. The
knowledge of the position of a qubit lost is assumed for the
erasure code to work, but this can be satisfied again with the use
of a probe mode. The probe mode can be used in a nondestructive
operation (a quantum nondemolition measurement) to determine if
the qubit is present or not without measuring its quantum state
\cite{Munro2005B}. There are a number of erasure codes we could
consider, but we will focus here on one presented recently by
Ralph {\it et al.} \cite{Ralph2005,Gilchrist2005} for linear
optical quantum computation. The logical basis states are defined
by tensor products of Bell states as
$\ket{0}_L^{(n)}=(\ket{00}+\ket{11})^{\otimes n}/2^{n/2}$ and
$\ket{1}_L^{(n)}=(\ket{01}+\ket{10})^{\otimes n}/2^{n/2}$. Here
$n$ is the number of Bell states used in the encoding. An
arbitrary state can then be written as
$c_0\ket{0}_L^{(n)}+c_1\ket{1}_L^{(n)}$ and can be created from
$c_0\ket{0}+c_1\ket{1}$ using only local operations and parity
gates. For simplicity, consider the $n=2$ (4 qubits) code where
qubit 1 is prepared as $c_0\ket{0}+c_1\ket{1}$ with the remaining
three in the state $\ket{0}$. The encoding procedure begins by
performing a Hadamard transformation on qubit 3 and then applying
the parity gate between qubits 1 and 3. This results in the state
$c_0\ket{0000}+c_1\ket{1010}$, after a bit-flip operation if the
odd-parity state occurred. We then apply the symmetrizer gate
between qubits 1/2 and 3/4. This transforms the state to $c_0
(\ket{00}+\ket{11})(
\ket{00}+\ket{11})/2+c_1(\ket{01}+\ket{10})(\ket{01}+\ket{10})/2$,
which is our encoded state
$c_0\ket{0}_L^{(2)}+c_1\ket{1}_L^{(2)}$. Larger $n$ states can be
prepared in a similar fashion.  To create the encoded state using
$n$ Bell pairs $c_0\ket{0}_L^{(n)}+c_1\ket{1}_L^{(n)}$ starting
with $c_0\ket{0}_L^{(n-1)}+c_1\ket{1}_L^{(n-1)}$, we perform
measurement on the second qubit of one of remaining Bell pairs in
the $\ket{0}$/$\ket{1}$ basis. If the measurement result is
$\ket{0}$, the state is projected onto
\begin{eqnarray}\label{parity-eqn-n}
c_0\ket{0}_L^{(n-2)} \ket{00}+c_1\ket{1}_L^{(n-2)}\ket{10}.
\end{eqnarray}
A similar state is obtained if the measurement result is
$\ket{1}$, which can be transformed via local operations to
Eq.~(\ref{parity-eqn-n}). By adding two qubits prepared in
$\ket{0}$ and applying the symmetrizer gates, we obtain
$c_0\ket{0}_L^{(n)}+c_1\ket{1}_L^{(n)}$.

Now consider that a qubit is lost and we have identified its
location. That causes our encoded state to become mixed, but we
can easily rectify this. We measure in the $\ket{+}/ \ket{-}$
basis the remaining qubit of the Bell pair in which a qubit is
lost. We then perform a phase-flip operation if the $\ket{-}$
state is measured. The effect of this operation is to transform
the encoded state according to $
c_0\ket{0}_L^{(n)}+c_1\ket{1}_L^{(n)}\rightarrow
c_0\ket{0}_L^{(n-1)}+c_1\ket{1}_L^{(n-1)}$. We have lost one Bell
pair (2 qubits) worth of encoding, but successfully regained the
original encoded information. To restore the fully encoded state
$c_0\ket{0}_L^{(n)}+c_1\ket{1}_L^{(n)}$, we further perform the
last step of the encoding procedure described above.

We now analyze error propagation during the error correcting
procedures.  Consider first the conventional error correcting
scheme using ancilla qubits. Suppose we use the 3-qubit bit-flip
code and measure $Z_1Z_2$ using an ancilla qubit at an error rate
$\epsilon$ (due to storage error and gate error) and two
controlled-NOT gates to the ancilla qubit conditioned on qubit 1
and 2, respectively. Then an error on the ancilla qubit propagates
to both qubits 1 and 2, inducing two errors in the data qubits
with $\epsilon$. Our proposed error correcting scheme, the parity
gate is the fundamental component.  In this gate, there are two
types of errors on the probe mode that propagate back to the data
qubits: (1) an intrinsic measurement error and (2) errors due to
loss, decoherence or noise on the probe mode.

The first type of errors arises from the fact that the states
$\ket{\alpha}$ and $\ket{\alpha e^{\pm i\theta}}$ of the probe
mode are not orthogonal and a measurement result in one parity
subspace could have come form the opposite parity state. This
intrinsic error, given by $P_{\rm err}(\theta)={\rm
erfc}[|\alpha|\sin\theta/\sqrt{2}]/2$, results in a wrong
error syndrome, which may introduce a bit-flip error in
the error correction procedure. This can be suppressed
(made small) when $|\alpha|\theta\gg 1$ \cite{Nemoto2004}. For
instance with $|\alpha|\theta=3.09$, $P_{\rm err}\sim 10^{-3}$.

The second type of errors includes the photon loss due to decoherence.
This causes dephasing, corresponding to phase flip errors, in the
original two-qubit state\cite{Munro2005A}. The degree of dephasing
is characterized by the parameter $\gamma = \eta^2\alpha|^2 \theta^2/2$
where $\eta^2$ is the percentage of photons lost from the probe mode as it
propogates through the parity gate. $\gamma $ must be keep small for
the dephasing to have a neligible effect. This in effect requires
$\eta \ll 1/ |\alpha|\theta$ which can be simply satisfied as long
$1\ll |\alpha|\theta<10$. For instance with $|\alpha|\theta=3.09$
and $\eta=0.035$. the error due to dephasing is of the order
$10^{-3}$. Now as we make $|\alpha|\theta$ larger, $\eta$ needs to decrease
to keep the dephasing error small. Other potential errors worth mentioning
are associated with differences in $\theta$ between the various qubits and
uncertainty in the value of $\theta$. Both of these errors can be managed
and are small when $\Delta\theta/\theta\ll 1$.

To summarize we have presented the error correcting procedures and
circuits based on weak nonlinearity between qubits and robust
continuous variable probe modes. Our error correcting procedures
have several distinct differences over the conventional method of
quantum error correction
\cite{Shor1995,Steane1996,Calderbank1996,Shor1996}. First, as
fragile ancilla qubits are replaced by robust continuous variable
probe modes, costly preparation of ancilla qubits is substituted
by the easy preparation of the probe modes. The use of such probe
modes also gives us freedom in constructing quantum circuits since
direct interaction between qubits is not necessary.

Secondly, we have also shown that in general the error correcting
circuits can be used for encoding an unknown state as well with a
slight modification. This property is general hence applicable to
the standard error-correction encoding procedure. Such a way of
encoding might have an advantage even with ancilla qubits where
ancilla qubits are used with probe modes or the use of
controlled-NOT gates on qubits are restricted. The easy
preparation and initialization of probe mode together with the
efficient parity gates make the error detection and correction
procedure significantly simpler than the standard error correction
scheme. Our new approach can be applied to the existing error
correcting codes including the 3-qubit bit-flip (phase-flip) code,
Shor and erasure codes and so has wide applicability in many
physical implementation of quantum computation ranging from the
solid-state to optics.

\noindent {\em Acknowledgments}: We would like to thank I.
L.Chuang, T. D. Ladd, R. Laflamme, P. van Loock, R. Van Meter and
Y. Yamamoto for valuable discussions. This work was supported in
part by JST SORST, JSPS, MIC, QAP, and Asahi-Glass research
grants.

\bibliography{error_code_fy25}

\begin{thebibliography}{19}
\expandafter\ifx\csname natexlab\endcsname\relax\def\natexlab#1{#1}\fi
\expandafter\ifx\csname bibnamefont\endcsname\relax
  \def\bibnamefont#1{#1}\fi
\expandafter\ifx\csname bibfnamefont\endcsname\relax
  \def\bibfnamefont#1{#1}\fi
\expandafter\ifx\csname citenamefont\endcsname\relax
  \def\citenamefont#1{#1}\fi
\expandafter\ifx\csname url\endcsname\relax
  \def\url#1{\texttt{#1}}\fi
\expandafter\ifx\csname urlprefix\endcsname\relax\def\urlprefix{URL }\fi
\providecommand{\bibinfo}[2]{#2}
\providecommand{\eprint}[2][]{\url{#2}}

\bibitem[{\citenamefont{Nielsen and Chuang}(2000)}]{Nielsen2000}
\bibinfo{author}{\bibfnamefont{M.~A.} \bibnamefont{Nielsen}} \bibnamefont{and}
  \bibinfo{author}{\bibfnamefont{I.~L.} \bibnamefont{Chuang}},
  \emph{\bibinfo{title}{Quantum Computation and Quantum Information}}
  (\bibinfo{publisher}{Cambridge University Press}, \bibinfo{year}{2000}).

\bibitem[{\citenamefont{Lo et~al.}(1998)\citenamefont{Lo, Popescu, and
  Spiller}}]{lo98}
\bibinfo{editor}{\bibfnamefont{H.-K.} \bibnamefont{Lo}},
  \bibinfo{editor}{\bibfnamefont{S.}~\bibnamefont{Popescu}}, \bibnamefont{and}
  \bibinfo{editor}{\bibfnamefont{T.~P.} \bibnamefont{Spiller}}, eds.,
  \emph{\bibinfo{title}{Introduction to Quantum Computation and Information}}
  (\bibinfo{publisher}{World Scientific Publishing}, \bibinfo{year}{1998}).

\bibitem[{\citenamefont{Gisin et~al.}(2002)\citenamefont{Gisin, Ribordy,
  Tittel, and Zbinden}}]{gis02}
\bibinfo{author}{\bibfnamefont{N.}~\bibnamefont{Gisin}},
  \bibinfo{author}{\bibfnamefont{G.}~\bibnamefont{Ribordy}},
  \bibinfo{author}{\bibfnamefont{W.}~\bibnamefont{Tittel}}, \bibnamefont{and}
  \bibinfo{author}{\bibfnamefont{H.}~\bibnamefont{Zbinden}},
  \bibinfo{journal}{Rev. Mod. Phys.} \textbf{\bibinfo{volume}{74}},
  \bibinfo{pages}{145} (\bibinfo{year}{2002}).

\bibitem[{\citenamefont{Spiller et~al.}(2005)\citenamefont{Spiller, Munro,
  Barrett, and Kok}}]{spi05}
\bibinfo{author}{\bibfnamefont{T.~P.} \bibnamefont{Spiller}},
  \bibinfo{author}{\bibfnamefont{W.~J.} \bibnamefont{Munro}},
  \bibinfo{author}{\bibfnamefont{S.~D.} \bibnamefont{Barrett}},
  \bibnamefont{and} \bibinfo{author}{\bibfnamefont{P.}~\bibnamefont{Kok}},
  \bibinfo{journal}{Contemporary Physics} \textbf{\bibinfo{volume}{46}},
  \bibinfo{pages}{407} (\bibinfo{year}{2005}).

\bibitem[{\citenamefont{Shor}(1995)}]{Shor1995}
\bibinfo{author}{\bibfnamefont{P.~W.} \bibnamefont{Shor}},
  \bibinfo{journal}{Phys. Rev. A} \textbf{\bibinfo{volume}{52}},
  \bibinfo{pages}{R2493} (\bibinfo{year}{1995}).

\bibitem[{\citenamefont{Steane}(1996)}]{Steane1996}
\bibinfo{author}{\bibfnamefont{A.}~\bibnamefont{Steane}},
  \bibinfo{journal}{Proc. Roy. Soc. London Ser. A}
  \textbf{\bibinfo{volume}{452}}, \bibinfo{pages}{2551} (\bibinfo{year}{1996}).

\bibitem[{\citenamefont{Calderbank and Shor}(1996)}]{Calderbank1996}
\bibinfo{author}{\bibfnamefont{A.~R.} \bibnamefont{Calderbank}}
  \bibnamefont{and} \bibinfo{author}{\bibfnamefont{P.~W.} \bibnamefont{Shor}},
  \bibinfo{journal}{Phys. Rev. A} \textbf{\bibinfo{volume}{54}},
  \bibinfo{pages}{1098} (\bibinfo{year}{1996}).

\bibitem[{\citenamefont{Shor}(1996)}]{Shor1996}
\bibinfo{author}{\bibfnamefont{P.~W.} \bibnamefont{Shor}}, in
  \emph{\bibinfo{booktitle}{Proceedings of 37th Annual Symposium on Foundations
  of Computer Science}} (\bibinfo{publisher}{IEEE Comput. Soc. Press},
  \bibinfo{year}{1996}), p.~\bibinfo{pages}{56}.

\bibitem[{\citenamefont{{T. P. Spiller {\it et al.}}}(2006)}]{Spiller2005}
\bibinfo{author}{\bibnamefont{{T. P. Spiller {\it et al.}}}},
  \bibinfo{journal}{New J. Phys.} \textbf{\bibinfo{volume}{8}},
  \bibinfo{pages}{30} (\bibinfo{year}{2006}).

\bibitem[{\citenamefont{Santori et~al.}(2002)\citenamefont{Santori, Fattal,
  Vuckovic, Solomon, and Yamamoto}}]{Santori2002}
\bibinfo{author}{\bibfnamefont{C.}~\bibnamefont{Santori}},
  \bibinfo{author}{\bibfnamefont{D.}~\bibnamefont{Fattal}},
  \bibinfo{author}{\bibfnamefont{J.}~\bibnamefont{Vuckovic}},
  \bibinfo{author}{\bibfnamefont{G.~S.} \bibnamefont{Solomon}},
  \bibnamefont{and} \bibinfo{author}{\bibfnamefont{Y.}~\bibnamefont{Yamamoto}},
  \bibinfo{journal}{Nature} \textbf{\bibinfo{volume}{419}},
  \bibinfo{pages}{594} (\bibinfo{year}{2002}).

\bibitem[{\citenamefont{{J. P. Reithmaier {\it et
  al.}}}(2004)}]{Reithmaier2004}
\bibinfo{author}{\bibnamefont{{J. P. Reithmaier {\it et al.}}}},
  \bibinfo{journal}{Nature} \textbf{\bibinfo{volume}{432}},
  \bibinfo{pages}{197} (\bibinfo{year}{2004}).

\bibitem[{\citenamefont{{T. Yoshie {\it et al.}}}(2004)}]{Yoshie2004}
\bibinfo{author}{\bibnamefont{{T. Yoshie {\it et al.}}}},
  \bibinfo{journal}{Nature} \textbf{\bibinfo{volume}{432}},
  \bibinfo{pages}{200} (\bibinfo{year}{2004}).

\bibitem[{\citenamefont{Akahane et~al.}(2003)\citenamefont{Akahane, Asano,
  Song, and Noda}}]{Akahane2003}
\bibinfo{author}{\bibfnamefont{Y.}~\bibnamefont{Akahane}},
  \bibinfo{author}{\bibfnamefont{T.}~\bibnamefont{Asano}},
  \bibinfo{author}{\bibfnamefont{B.-S.} \bibnamefont{Song}}, \bibnamefont{and}
  \bibinfo{author}{\bibfnamefont{S.}~\bibnamefont{Noda}},
  \bibinfo{journal}{Nature} \textbf{\bibinfo{volume}{425}},
  \bibinfo{pages}{944} (\bibinfo{year}{2003}).

\bibitem[{\citenamefont{Nemoto and Munro}(2004)}]{Nemoto2004}
\bibinfo{author}{\bibfnamefont{K.}~\bibnamefont{Nemoto}} \bibnamefont{and}
  \bibinfo{author}{\bibfnamefont{W.~J.} \bibnamefont{Munro}},
  \bibinfo{journal}{Phys. Rev. Lett.} \textbf{\bibinfo{volume}{93}},
  \bibinfo{pages}{250502} (\bibinfo{year}{2004}).

\bibitem[{\citenamefont{Munro et~al.}(2005{\natexlab{a}})\citenamefont{Munro,
  Nemoto, and Spiller}}]{Munro2005A}
\bibinfo{author}{\bibfnamefont{W.~J.} \bibnamefont{Munro}},
  \bibinfo{author}{\bibfnamefont{K.}~\bibnamefont{Nemoto}}, \bibnamefont{and}
  \bibinfo{author}{\bibfnamefont{T.~P.} \bibnamefont{Spiller}},
  \bibinfo{journal}{New Journal of Physics} \textbf{\bibinfo{volume}{7}},
  \bibinfo{pages}{137} (\bibinfo{year}{2005}{\natexlab{a}}).

\bibitem[{\citenamefont{Munro et~al.}(2005{\natexlab{b}})\citenamefont{Munro,
  Nemoto, Beausoleil, and Spiller}}]{Munro2005B}
\bibinfo{author}{\bibfnamefont{W.~J.} \bibnamefont{Munro}},
  \bibinfo{author}{\bibfnamefont{K.}~\bibnamefont{Nemoto}},
  \bibinfo{author}{\bibfnamefont{R.~G.} \bibnamefont{Beausoleil}},
  \bibnamefont{and} \bibinfo{author}{\bibfnamefont{T.~P.}
  \bibnamefont{Spiller}}, \bibinfo{journal}{Phys. Rev. A}
  \textbf{\bibinfo{volume}{71}}, \bibinfo{pages}{033819}
  (\bibinfo{year}{2005}{\natexlab{b}}).

\bibitem[{\citenamefont{Ralph et~al.}(2005)\citenamefont{Ralph, Hayes, and
  Gilchrist}}]{Ralph2005}
\bibinfo{author}{\bibfnamefont{T.~C.} \bibnamefont{Ralph}},
  \bibinfo{author}{\bibfnamefont{A.~J.~F.} \bibnamefont{Hayes}},
  \bibnamefont{and}
  \bibinfo{author}{\bibfnamefont{A.}~\bibnamefont{Gilchrist}},
  \bibinfo{journal}{Phys. Rev. Lett.} \textbf{\bibinfo{volume}{95}},
  \bibinfo{pages}{100501} (\bibinfo{year}{2005}).

\bibitem[{\citenamefont{Gilchrist et~al.}(2005)\citenamefont{Gilchrist, Hayes,
  and Ralph}}]{Gilchrist2005}
\bibinfo{author}{\bibfnamefont{A.}~\bibnamefont{Gilchrist}},
  \bibinfo{author}{\bibfnamefont{A.~J.~F.} \bibnamefont{Hayes}},
  \bibnamefont{and} \bibinfo{author}{\bibfnamefont{T.~C.} \bibnamefont{Ralph}},
  \bibinfo{journal}{quant-ph/0505125}  (\bibinfo{year}{2005}).

\bibitem[{\citenamefont{Nemoto and Munro}(2005)}]{Nemoto2005}
\bibinfo{author}{\bibfnamefont{K.}~\bibnamefont{Nemoto}} \bibnamefont{and}
  \bibinfo{author}{\bibfnamefont{W.~J.} \bibnamefont{Munro}},
  \bibinfo{journal}{Phys. Lett A} \textbf{\bibinfo{volume}{344}},
  \bibinfo{pages}{104} (\bibinfo{year}{2005}).

\end{thebibliography}
\end{document}